\documentclass[preprint,a4paper]{aastex63}

\usepackage{amsmath,amssymb}
\usepackage{natbib}
\usepackage{graphicx}
\usepackage{upgreek}
\usepackage{url}

\received{February 25, 2020}
\accepted{May 11, 2020}
\submitjournal{ApJS}

\shorttitle{100 yr Solar Filament Dataset}
\shortauthors{Lin et al.}

\begin{document}

\title{A New Comprehensive Data Set of Solar Filaments of 100 yr Interval. I.}

\correspondingauthor{GangHua Lin}
\email{lgh@nao.cas.cn}

\author{GangHua Lin}
\affiliation{CAS Key Laboratory of Solar Activity, National Astronomical Observatories, Chinese Academy of Sciences, Beijing 100101, P.R. China}

\author{GaoFei Zhu}
\affiliation{CAS Key Laboratory of Solar Activity, National Astronomical Observatories, Chinese Academy of Sciences, Beijing 100101, P.R. China}
\affiliation{School of Astronomy and Space Science, University of Chinese Academy of Sciences, Beijing 100049, P.R. China}
\affiliation{University of Chinese Academy of Sciences, Beijing 100049, P.R. China}

\author[0000-0003-1675-1995]{Xiao Yang}
\affiliation{CAS Key Laboratory of Solar Activity, National Astronomical Observatories, Chinese Academy of Sciences, Beijing 100101, P.R. China}

\author[0000-0002-9961-4357]{YongLiang Song}
\affiliation{CAS Key Laboratory of Solar Activity, National Astronomical Observatories, Chinese Academy of Sciences, Beijing 100101, P.R. China}

\author{Mei Zhang}
\affiliation{CAS Key Laboratory of Solar Activity, National Astronomical Observatories, Chinese Academy of Sciences, Beijing 100101, P.R. China}
\affiliation{School of Astronomy and Space Science, University of Chinese Academy of Sciences, Beijing 100049, P.R. China}

\author[0000-0002-1396-7603]{Suo Liu}
\affiliation{CAS Key Laboratory of Solar Activity, National Astronomical Observatories, Chinese Academy of Sciences, Beijing 100101, P.R. China}
\affiliation{School of Astronomy and Space Science, University of Chinese Academy of Sciences, Beijing 100049, P.R. China}

\author{XiaoFan Wang}
\affiliation{CAS Key Laboratory of Solar Activity, National Astronomical Observatories, Chinese Academy of Sciences, Beijing 100101, P.R. China}

\author{JiangTao Su}
\affiliation{CAS Key Laboratory of Solar Activity, National Astronomical Observatories, Chinese Academy of Sciences, Beijing 100101, P.R. China}
\affiliation{School of Astronomy and Space Science, University of Chinese Academy of Sciences, Beijing 100049, P.R. China}

\author{Sheng Zheng}
\affiliation{College of Science, China Three Gorges University, Yichang 443002, P.R. China}

\author{JiaFeng Zhang}
\affiliation{University of Chinese Academy of Sciences, Beijing 100049, P.R. China}
\affiliation{College of Science, China Three Gorges University, Yichang 443002, P.R. China}

\author{DongYi Tao}
\affiliation{College of Science, China Three Gorges University, Yichang 443002, P.R. China}

\author{ShuGuang Zeng}
\affiliation{College of Science, China Three Gorges University, Yichang 443002, P.R. China}

\author[0000-0002-5233-565X]{HaiMin Wang}
\affiliation{Institute for Space Weather Sciences, New Jersey Institute of Technology, Newark, NJ 07102-1982, USA}

\author{Chang Liu}
\affiliation{Institute for Space Weather Sciences, New Jersey Institute of Technology, Newark, NJ 07102-1982, USA}

\author{Yan Xu}
\affiliation{Institute for Space Weather Sciences, New Jersey Institute of Technology, Newark, NJ 07102-1982, USA}

\begin{abstract}

Filaments are very common physical phenomena on the Sun and are often taken as important proxies of solar magnetic activities. The study of filaments has become a hot topic in the space weather research. For a more comprehensive understanding of filaments, especially for an understanding of solar activities of multiple solar cycles, it is necessary to perform a combined multifeature analysis by constructing a data set of multiple solar cycle data. To achieve this goal, we constructed a centennial data set that covers the H$\alpha$ data from five observatories around the world. During the data set construction, we encountered varieties of problems, such as data fusion, accurate determination of the solar edge, classifying data by quality, dynamic threshold, and so on, which arose mainly due to multiple sources and a large time span of data. But fortunately, these problems were well solved. The data set includes seven types of data products and eight types of feature parameters with which we can implement the functions of data searching and statistical analyses. It has the characteristics of better continuity and highly complementary to space observation data, especially in the wavelengths not covered by space observations, and covers many solar cycles (including more than 60 yr of high-cadence data). We expect that this new comprehensive data set as well as the tools will help researchers to significantly speed up their search for features or events of interest, for either statistical or case study purposes, and possibly help them get a better and more comprehensive understanding of solar filament mechanisms.

\end{abstract}


\section{Introduction}
\label{sec:intro}

Solar filaments, observed in the H$\alpha$ line of the solar chromosphere, are one of the basic indices of solar activity \citep{dAzambuja1923}.  They are often observed to be associated with flares and coronal mass ejections (CMEs) \citep{Golub1997,Galsgaard1999}, which are the major driving sources of hazardous space weather. Increasing observational evidences have suggested that filament eruptions, flares, and CMEs are different manifestations of one physical process at different evolutionary stages \citep{Gilbert2000,Gopalswamy2003}. While on one hand, in order to gain a better understanding of CMEs and advance our forecast capabilities, it is essential to identify early manifestations of CMEs such as filament eruptions and flares \citep{JingJ2004}, on the other hand, multiple cycles or long-term filament data can actually give us more information than just for CME studies. For example, the long-term variation of the statistical properties of filaments carries important information about the magnetic signature of the solar cycle, the polar crown filament demonstrates the interface between new and old solar cycle, and the possible changes of efficiency in the dynamo may also be probed by the study of filaments.

Specifically, the above problems can be understood through the changes of the features or parameters of solar filaments. The solar filament is a dark strip projected onto the surface of the Sun. It is the dividing line of the local magnetic field with opposite polarity and reflects the complex structure of the solar magnetic field to some extent. The main structure of the filament is spine and foots. Bar is a branch of the filament. The spine and foots of filament are dynamic \citep{Martin2008}. There are certain rules for the generation, development and disappearance of filaments, but they may suddenly become active in a very short period of time, or even disappear suddenly, and its disappearance is divided into different types. The length of filament is typically many thousands kilometers. It can be longer than the radius of the Sun or just a few mega meters. A filament rapidly forms over a period of a day but then typically persists for several weeks and, in some cases, several months. These complex structures and dynamic characteristics of filaments are closely related to the magnetic field on the Sun. The process of evolution and formation involves the recombination and transformation of the magnetic field on the Sun into a filament magnetic field under certain conditions. However, it is not clear how the magnetic field on the Sun is converted to a filament magnetic field nor is the detailed magnetic field structure of the filament. The study of the formation of the filament not only helps us to understand the process of recombination of magnetic fields on the Sun into a filament magnetic field but it also helps us to understand the support of filament and the detailed structure of its magnetic field.

However, so far, a database of continuous and long-term filament observations is not in place. Therefore, data containing long (multiple solar cycles) and continuous filament observations need to be collected, and a dataset of them needs to be constructed and the filament parameters that involve the formation, disappearance, dynamic process, structure, and shape need to be extracted.  In order to achieve this goal, the Comprehensive Data set of One-Hundred-Year Solar Filament (CDOHSF, \url{http://sun.bao.ac.cn/solarfilament/}) has been developed for the international scientific community.

In Section~\ref{sec:data}, we describe the sources of the data that we collected. In Section~\ref{sec:dataProc}, the processing flow to these original data is introduced. In Section~\ref{sec:dataProd}, the products and a query tool are introduced with examples. In the last section, conclusion and future work are presented.

\section{Data Sources}
\label{sec:data}

There are already a few filament catalogs or databases exist in the community. Here we give a brief list of some of them.

1. \url{http://www.bbso.njit.edu/Research/FDHA/}  is an online FTP filament archive, provided by the Big Bear Solar Observatory (BBSO), USA. The data list is from 1982 to today.

2. \url{http://ftp.kso.ac.at} is an online FTP filament archive of the Kanzelh\"{o}he Solar Observatory (KSO), Austria. The data list is from 1997. On the FTP site, there are some information about the catalog structure of the online data files, FITS header keywords, and so on.

3. \url{https://kso.iiap.res.in/data} is an online filament data archive of the Kodaikanal Solar Observatory, India. There is an interactive data visualization tool that tells you how many observational images are there for each year on the webpage. Their chromospheric data is from 1907.

4. \url{http://sun.bao.ac.cn/hsos_data/full_disk/h-alpha/} is an online filament data archive of the Huairou Solar Observing Station (HSOS), China. The data list is from 2000.

5. \url{http://en.solarstation.ru/archive} is from the Kislovodsk Mountain Station, Russia. There are lots of tables that contain characteristics of solar filaments from 1959 to today.

6. \url{https://www.ngdc.noaa.gov/stp/space-weather/solar-data/solar-features/prominences-filaments/filaments/} is a site of the National Centers for Environmental Information (NOAA), USA. There are tables on the site address. These tables contain information on the coordinates, time, Carrington number, minimum and maximum  lengths of the filament during the solar disk passage, radial velocity, importance, characteristics of filament (stable, unstable, end, broad, clear, diffuse, and discontinuous), and so on. Their filaments cover two time intervals (1919 -- 1956 and 1957 -- 2000), with data from several observatories. The features are obtained from synoptic maps. There are tables of the limb feature from 2000 to 2009 and tables of filament disappearances from 1964 to 1980 on the website \citep{Wright1983,Mouradian1994,Golub1997,Galsgaard1999,Low2003}.

7. So far, \url{http://aia.cfa.harvard.edu/filament/} is the only site that provides interactive visualization tool. It is based on the Atmospheric Imaging Assembly (AIA) satellite observation data (20100518 -- 20140920) used to obtain the basic information and a large number of observational characters of filament eruption events --- for example, the start time of a filament, duration time of eruption, position $(X'', Y'')$, rating, type, flare association, grade of associated flare, and so on. Through buttons, one can watch movies and download the catalog, etc. A collection of the highest rated events is also available as a YouTube playlist.

\begin{figure}[!htbp]
    \centering
    \includegraphics[width=\textwidth]{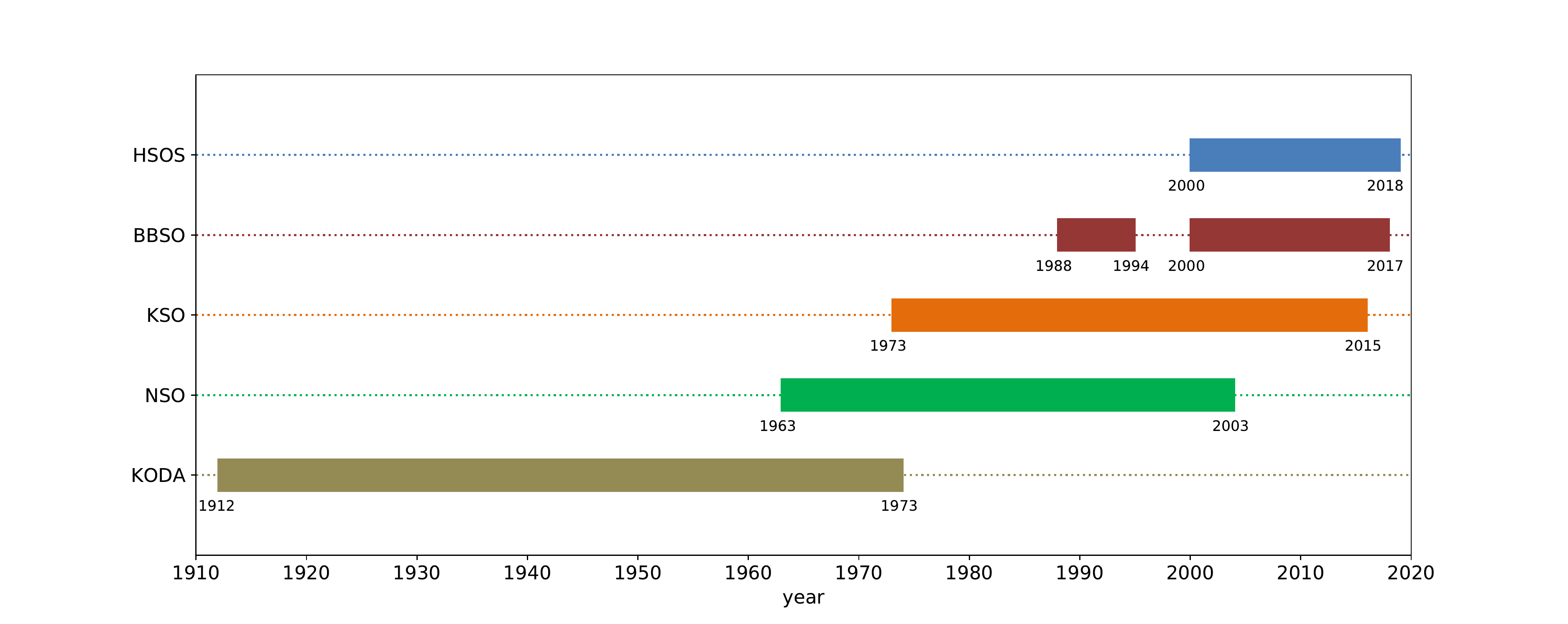}
    \caption{Year span of the data of the five observatories.}
    \label{fig:HaDataSpan}
\end{figure}

Among the H$\alpha$ data of the six observatories we surveyed, the data of the Kislovodsk Mountain Astronomical Station is still in the process of being digitized. Therefore, we collected the original data from five observatories around the world. They are listed in Table~\ref{tab:fileStat}. They are: the BBSO, National Solar Observatory of the USA (NSO), KSO, Kodaikanal Observatory of India (KODA in this paper), and HSOS. Data of earlier years were digitized from films. Part of the data were provided by Prof. Haimin Wang's team (data from BBSO, NSO, and KSO, $\sim 37.2$ TB in total after compression), part are from our Huairou Solar Observing Station (HSOS, $\sim 8.1$ TB), and a fraction of them are downloaded through the web (KODA, $\sim 680$ GB). The complete and up-to-date list of integrated data source can be found via \url{http://sun.bao.ac.cn/solarfilament/}. The time span of the data can be seen from Figure~\ref{fig:HaDataSpan}.

\begin{table}
\begin{center}
\caption{Statistics on the data files collected from the five observatories.}
\label{tab:fileStat}
\begin{tabular}{cccc}
    \hline
    Observatory & Year Span & Number of Files & Disk Size (Compressed)   \\
    \hline
    HSOS & 2000 -- 2018 & 1,088,914 & 8.1 TB   \\
    BBSO & 1988 -- 1994, 2000 -- 2017 & 454,923 & 2.1 TB   \\
    KSO & 1973 -- 2015 & 1,046,085 & 5.1 TB   \\
    NSO & 1963 -- 2003 & 7,299,490 & 30 TB   \\
    KODA & 1912 -- 1973 & 18,313 & 680 GB   \\
    \hline
\end{tabular}
\end{center}
\end{table}

\section{Processing Procedures for the Original Data}
\label{sec:dataProc}

\begin{figure}[!htbp]
    \centering
    \includegraphics[width=0.4\textwidth]{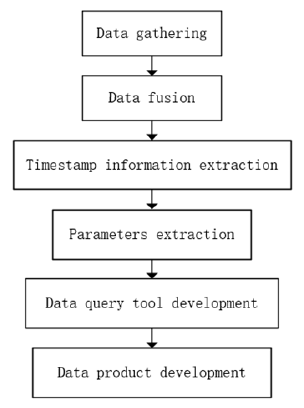}
    \caption{Flow of the data processing procedures.}
    \label{fig:dataProcFlow}
\end{figure}

A flow chart of the data processing procedures is shown in Figure~\ref{fig:dataProcFlow}. Each procedure will be introduced in the following subsections and sections, except that the procedure of ``data gathering'' has been introduced in the above section.

\subsection{Data Fusion}

\begin{figure}[!htbp]
    \centering
    \includegraphics[width=0.9\textwidth]{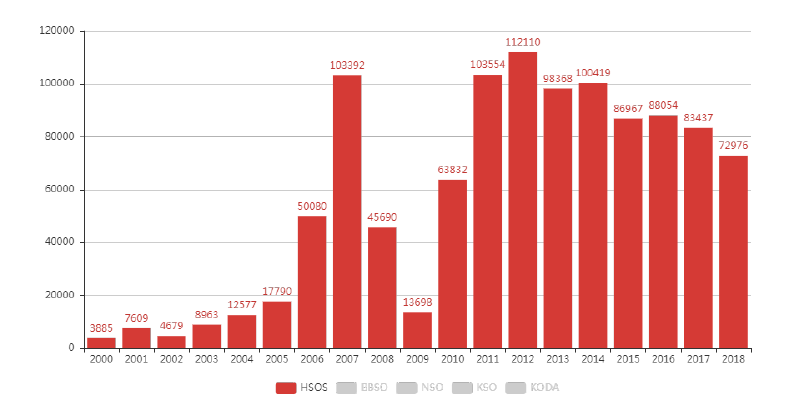}
    \caption{An example histogram of the data distribution of HSOS.}
    \label{fig:GHA-HSOS}
\end{figure}

We take the following steps to rearrange the raw data: we (1) copy data to uniformly named directories, (2) check original file names, (3) compress the files (gzip or pigz), (4) apply file name standardization, and (5) check level 0 file names. After these procedures, the total file (image) numbers and occupied disk sizes are listed in Table~\ref{tab:fileStat}. As shown in Figure~\ref{fig:HaDataSpan}, the time spans of the data in HSOS, BBSO, KSO, NSO, and KODA are 2000 -- 2018, 1988 -- 1994 and 2000 -- 2017, 1973 -- 2015, 1963 -- 2003, and 1912 -- 1973, respectively. Altogether these data consist of a dataset covering more than a hundred years or almost 10 solar cycles (SC15 -- SC24). A detailed data distribution for each observatory in years is available at \url{http://sun.bao.ac.cn/solarfilament/GHA_data_stat.html}, an example of which is shown in Figure~\ref{fig:GHA-HSOS}.

To guarantee consistent and smooth processing in the following steps, data standardization has been done at the very beginning. The work of data reorganization and standardization is time consuming but beneficial for the subsequent batch processing and analysis. File names and directory structure are standardized in the rearrangement of the raw data of all five observatories. Then when producing level 1 data from those raw level 0 data, nonuniform and limb darkening effects are removed; each image has been resized to put the solar disk in the image center with the same size in terms of pixel numbers; some images have also been rotated to make the north pole point upward.

The file names are formatted as ``$\langle$obs$\rangle$\_halph\_f[ildf]\_$\langle$yyyymmdd$\rangle$\_$\langle$hhmmss$\rangle$.fits.gz'', where $\langle$obs$\rangle$ is the name code of the observatory, ``f'' means full disk, ``i'' stands for initial images, ``l'' is for images with dark and flat field removed, ``d'' is for dark frames, and ``f'' is for flat field frames. ``$\langle$yyyymmdd$\rangle$'' and ``$\langle$hhmmss$\rangle$'' give the time information of the observation. This follows the file name convention used by BBSO and the Global H$\alpha$ Network Data Archive \footnote{\url{http://bbso.njit.edu/Archive/filenames.html}}. All the data are compressed through gzip or pigz to save storage space. And the directory tree is constructed as ``/$\langle$data\_root$\rangle$/$\langle$obs$\rangle$/$\langle$year$\rangle$/*.fits.gz''.

\begin{figure}[!htbp]
    \centering
    \includegraphics[width=\textwidth]{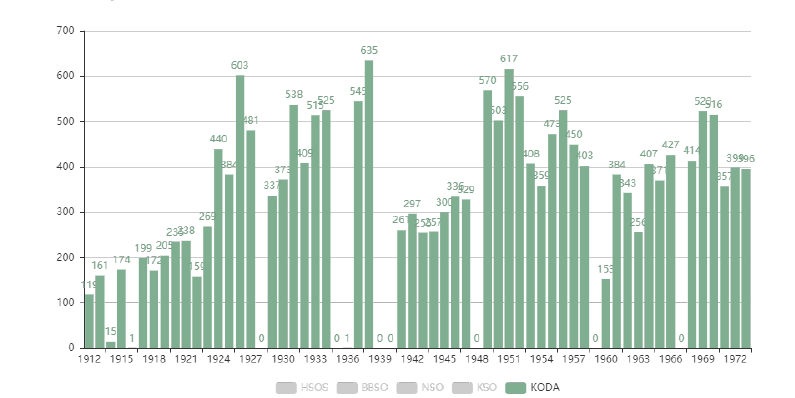}
    \caption{Collection of KODA observational data.}
    \label{fig:dataKODA}
\end{figure}

From Figure~\ref{fig:HaDataSpan}, we can see that the data in the early years is less dense. The data collection of KODA is shown in Figure~\ref{fig:dataKODA}. Data of NSO covers 1965 to 2003, for nearly 40 yr, and the quality of the data is high. This plays an important role in forming good continuity of the whole observation data. However, the observation time information of the NSO data was recorded on the observation image in the form of a time stamp. There were no time information in either the file names or the header of the FITS files. This leaves us with a tough task of getting the time information out in order to use these data for our data set. We will introduce our effort in Section~\ref{sec:timestamp}.

\subsection{Extraction of Time Stamp Information}
\label{sec:timestamp}

Extracting the time stamp information of NSO data is a huge project, because the number of the NSO images from 1963 to 2003 is almost 10 million. We have carried out the hard work of recognizing the time stamps in the scanned images and checking their correctness manually \citep{ZhangJF2019}. Then we rename all the NSO data accordingly, with the confirmed time recognized from the scanned time stamps.

\begin{figure}[!htbp]
    \centering
    \includegraphics[width=0.8\textwidth]{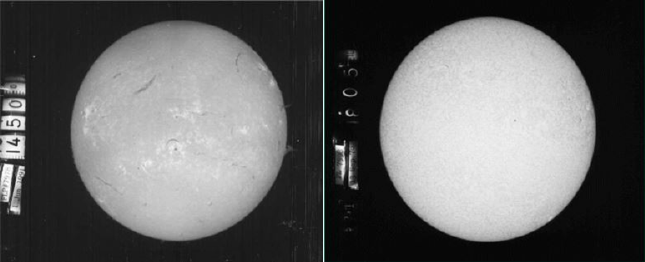}
    \caption{Two examples of scanned time stamps.}
    \label{fig:timestamp}
\end{figure}

The complex and changing style of the time stamps makes this task even more difficult. For example, there were different types of time stamp: black numbers with white background as shown in the left panel of Figure~\ref{fig:timestamp}, and white numbers with black background as shown in the right panel of Figure~\ref{fig:timestamp}. Sometimes the time digits can be blurred as the one shown in the right panel of Figure~\ref{fig:timestamp}, making it almost impossible to extract the needed information of the year, month, hour, minute, and second. There were also situations where some numbers were broken in the middle, or were superimposed with others, and so on.

\begin{figure}[!htbp]
    \centering
    \includegraphics[width=0.4\textwidth]{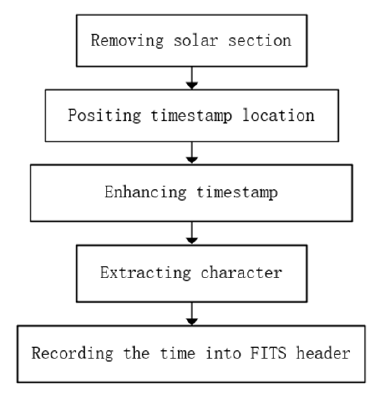}
    \caption{Flow chart of the time stamp information extraction.}
    \label{fig:timeExtFlow}
\end{figure}

The flow chart of the main procedures to do time stamp extraction is shown in Figure~\ref{fig:timeExtFlow}. First, the whole image is cut with only the time stamp being retained; then the time stamp is positioned and enhanced, characters in the time stamp are extracted, and these characters are then recognized.

When doing time character recognition, we first used a method of automatic identification (a convolutional neural network (CNN)) to read all the 8,699,681 time stamps; details of the identification method are described in \citet{ZhangJF2019}. Then more than 70 people spent 4.5 months doing manual verification for all these 8,699,681 time stamps. Among these time stamps, 7,760,911 images can be successfully identified by human eyes. These 7,760,911 time stamps are manually checked and corrected (if needed). Realizing that human checking may also make mistakes, we did a second round of manual checks to get an accuracy estimate. For each person, 1000 images they checked were randomly selected and rechecked by a second person. The result gives an average accuracy of $\sim 96$\% of all people.

\begin{figure}[!htbp]
    \centering
    \includegraphics[width=\textwidth]{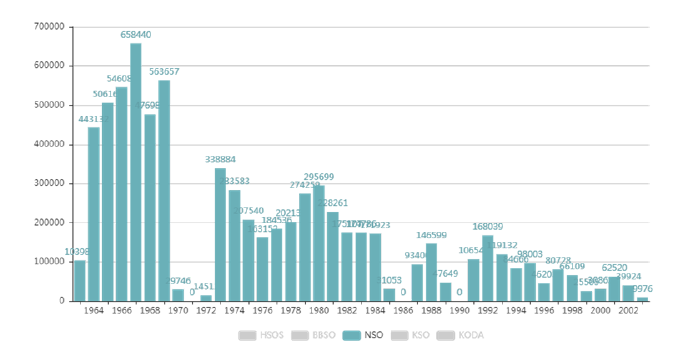}
    \caption{Statistics on NSO data.}
    \label{fig:dataNSO}
\end{figure}

Our effort revives the statistical usage of the historical NSO data. Now we can give statistical quantities of the NSO data whose time stamps have been successfully recognized as shown in Figure~\ref{fig:dataNSO},.

\subsection{The Physics Parameters Extraction}

\subsubsection{Preprocessing}

Since the collected H$\alpha$ images come from different periods and stations, some preprocessing steps need to be performed on these data.

\begin{figure}[!htbp]
    \centering
    \includegraphics[width=0.7\textwidth]{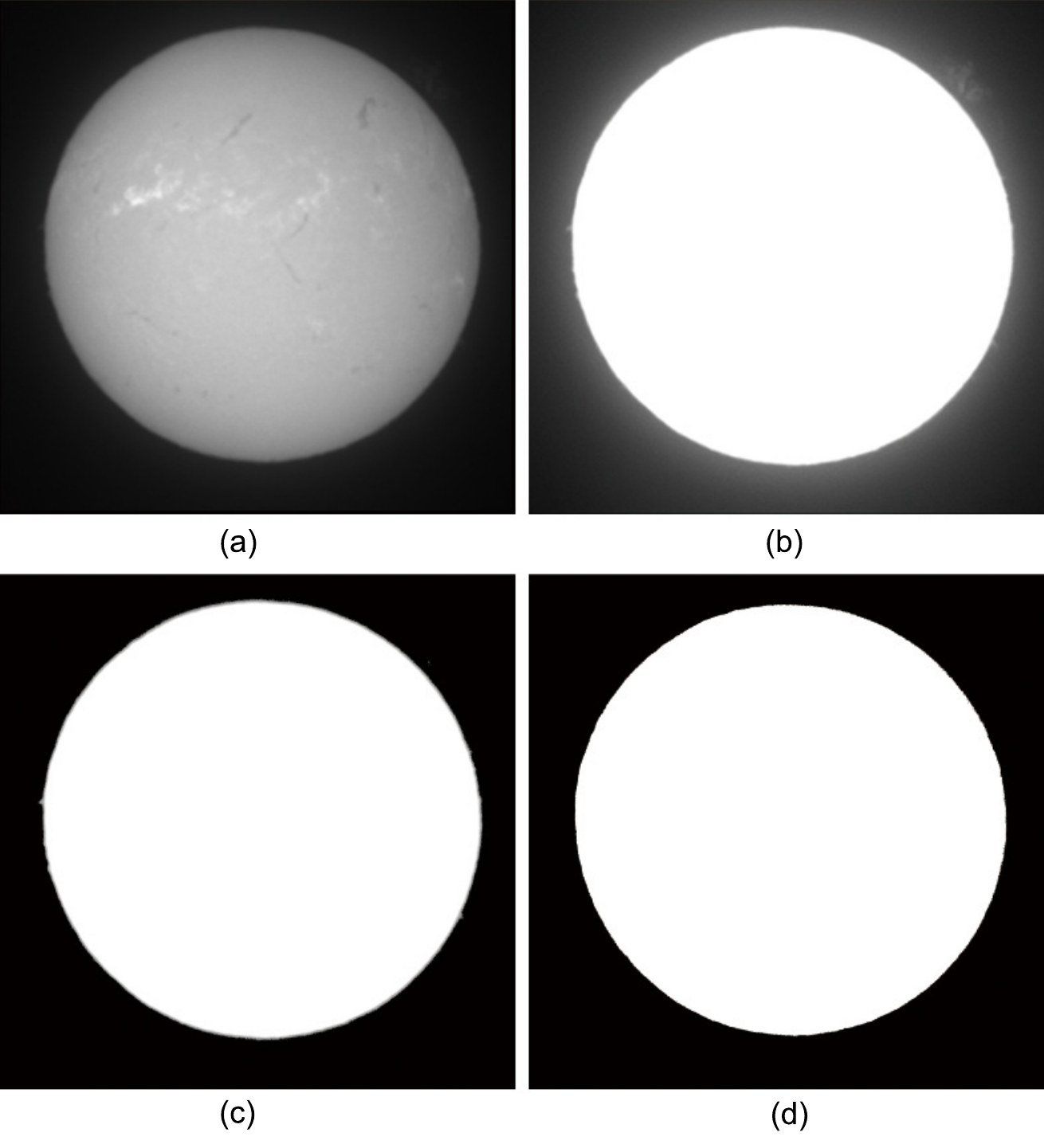}
    \caption{(a) An original H$\alpha$ full disk solar image. (b) Multiply the whole image by an integer to make the image maximum be 65,535. (c) For the image in panel (b), set all values lower than 32,768 to be zero to be used in step (2). (d) The resultant binary map, generated using the OTSU method in step (2).}
    \label{fig:circle}
\end{figure}

First, we use the cascading Hough circle detector \citep{YuanY2011} and the least squares circle fitting method to obtain the center and radius of the solar disk. In light of the existence of various noises in the historical data images, the Hough transform is a good way to crop the solar disk in the image because it is not sensitive to the noise. However, the calculation of the Hough transform is huge, so our approach is to apply the Hough transform to the resolution reduced image and get a rough center coordinate and the radius of the solar disk first. Then we use the more noise-sensitive but more computation-economical least squares circle fitting method to get the information about the image disk center and radius. The main steps, after roughly cropping the solar disk by using the Hough transform, are as follows:
\begin{enumerate}
  \item[(1)] We increase the contrast between the disk and the background. For each image, we first multiply the image by an integer. The integer is obtained by first dividing 65,535 by the maximum value of the image and then rounding the number up to an integer. Then we set all the values less than 32,768 to be zero. This modified image is then to be used in the next step (Figure~\ref{fig:circle}).
  \item[(2)] We use the OTSU algorithm \citep{Otsu1979} to segment out the solar disk from the image obtained in the previous step.
  \item[(3)] Edge detection is carried out on the previous segmentation result, and dilation operation and morphological thinning operation are carried out to get the edge detection result.
  \item[(4)] Based on the edge detection result, the number of pixels in all connected regions is counted, the largest connected region is reserved, and other regions are deleted to get the solar disk edge.
  \item[(5)] Based on the edge of the solar disk, the circularity `$m$' of it can be calculated using $m = 4\pi a/p^2$. If `$m$' is greater than the threshold value 0.9, the discrete points of the edge of the solar disk are fitted with the least square method, and the center and radius of the disk are calculated.
\end{enumerate}

Second, in the full disk H$\alpha$ image, the solar disk has decreasing brightness from the center to the limb, which is called limb darkening \citep{Ozhogina2009}. Limb darkening removal is necessary. Finally, after limb darkening removal, the background intensity of the solar disk is still nonuniform. We removed the uneven background intensity using the polynomial surface fitting method \citep{Bernasconi2005}.

\subsubsection{Filament Recognition}
\label{sec:filRec}

\begin{figure}[!htbp]
    \centering
    \includegraphics[width=0.7\textwidth]{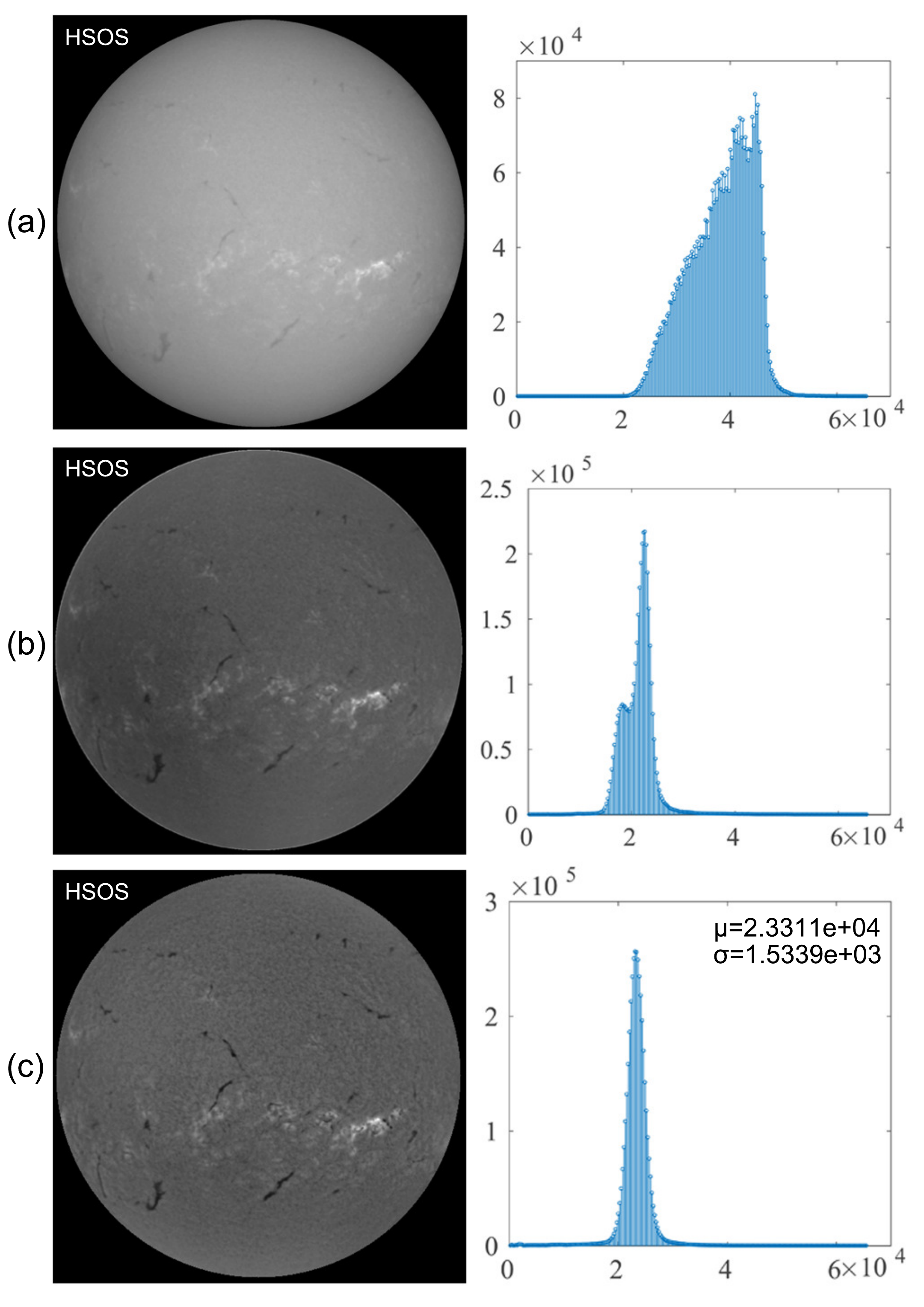}
    \caption{Top panels: the original image and its histogram. Middle panels: the image after removing limb darkening and its histogram. Bottom panels: the image after further removing the uneven background and its histogram, which now shows a normal distribution.}
    \label{fig:gauss}
\end{figure}

After the above preprocessing, we find that the intensity distribution on the solar disk is normal when we focus only on the inside of the solar disk (excluding the outside). In other words, when we count the number of each intensity value on the solar disk, the curve is in accordance with a normal distribution as shown in Figure~\ref{fig:gauss}. We use a single Gaussian function to fit the profile and get the values of the mean and standard deviation.

Since solar filaments present as dark features on the disk (the intensity value is small), the intensity threshold we want to get should be in the left portion of the normal distribution curve. According to the $3\sigma$ principle of normal distribution, we can get the corresponding intensity threshold. Its expression is
\begin{equation}
f(x) = \frac{1}{\sigma \sqrt{2\pi}}\mathrm{e}^{(\frac{x-\mu}{\sigma})^2} ~~~,
\end{equation}
where $\mu$ and $\sigma$ are obtained by fitting the Gaussian function. Then we use the following expression
\begin{equation}
x = \mu - 3\sigma
\end{equation}
to get the intensity threshold, and we apply it to the original binary image.

\begin{figure}[!htbp]
    \centering
    \includegraphics[width=0.7\textwidth]{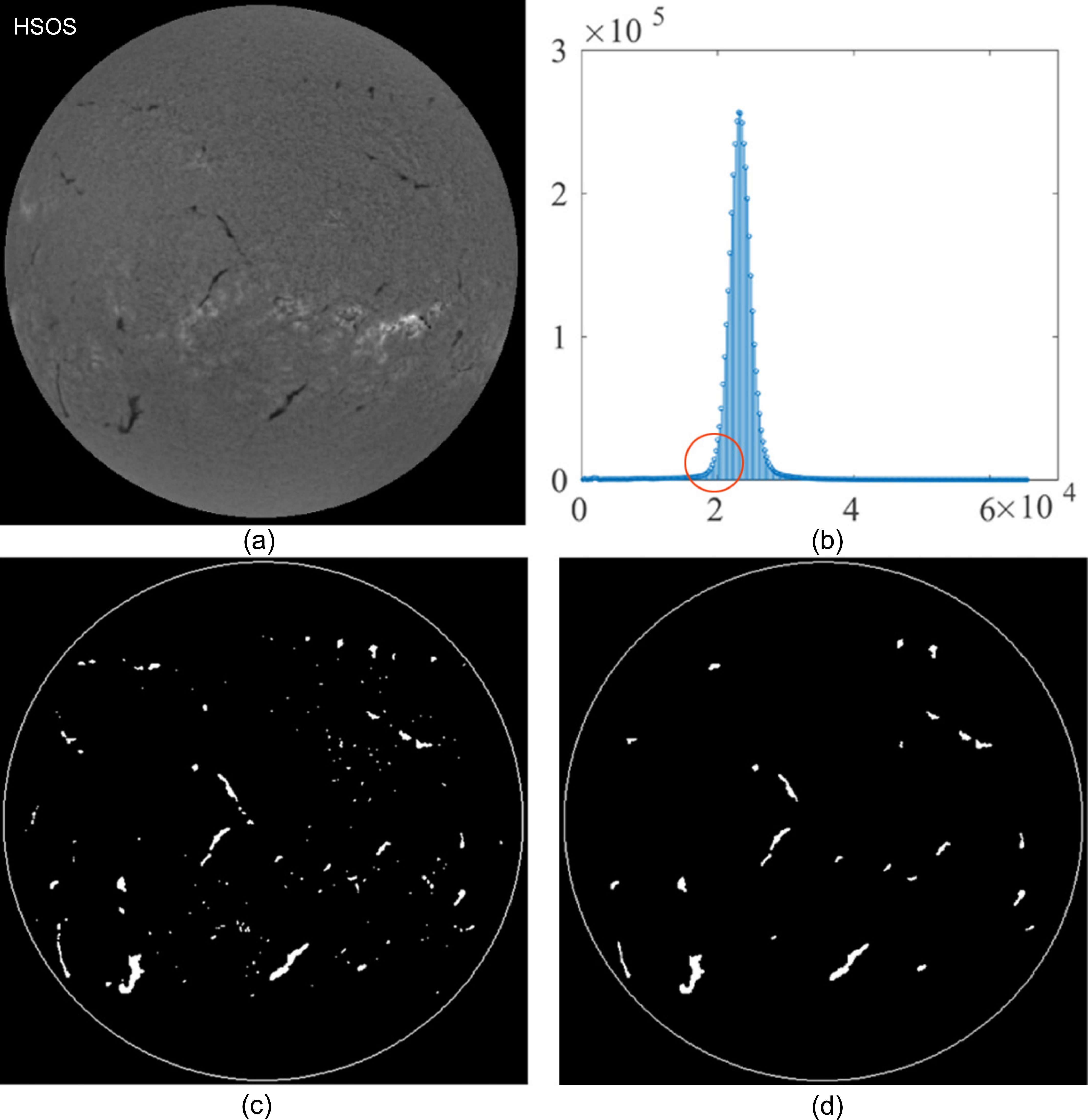}
    \caption{(a) The image after removing limb darkening and the uneven background. (b) Its histogram and the intensity threshold obtained (shown in red circle). (c) The result of segmentation. (d) Final segmentation result after removing small noise points.}
    \label{fig:filaExtResult}
\end{figure}

\begin{figure}[!htbp]
    \centering
    \includegraphics[width=0.8\textwidth]{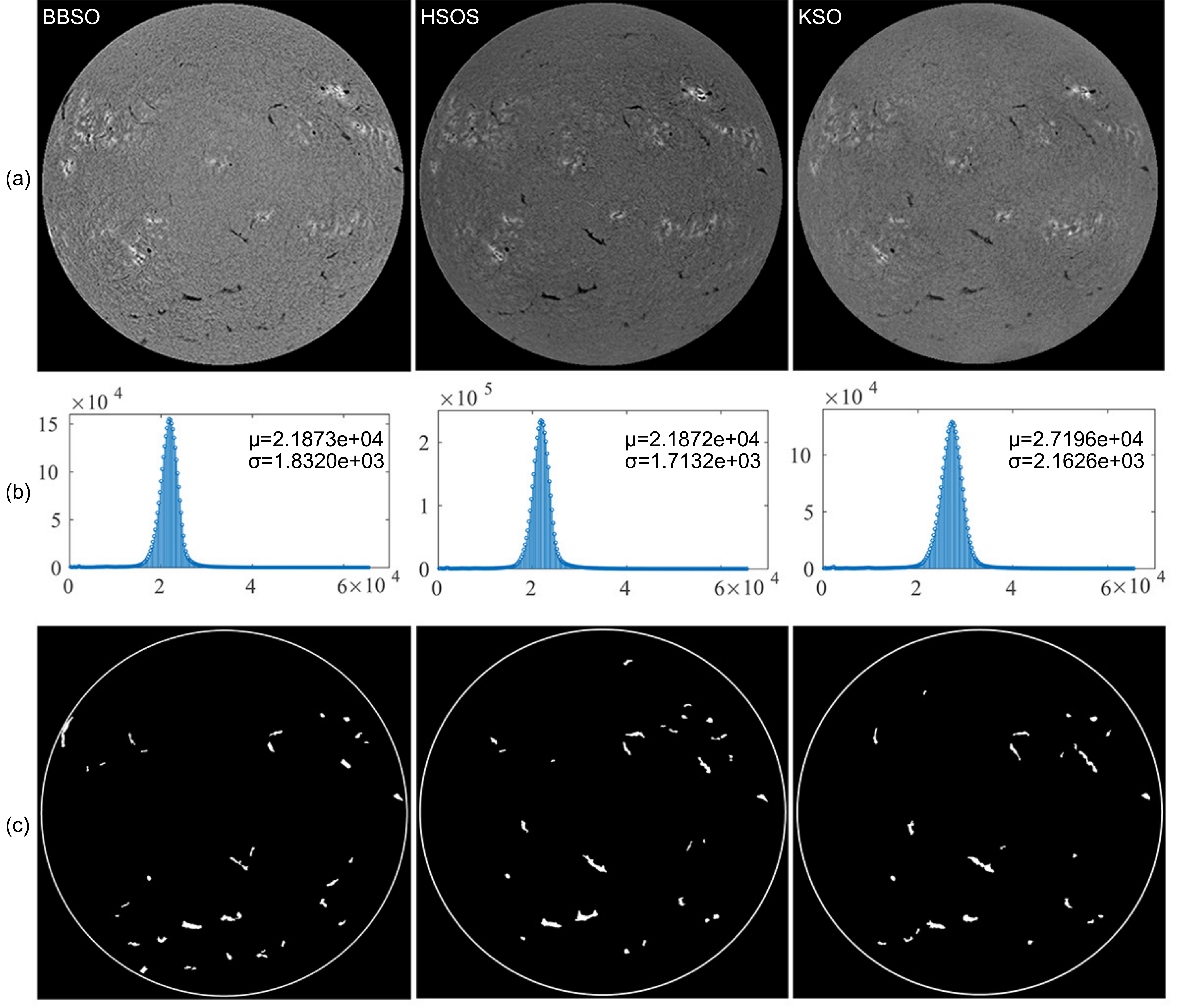}
    \caption{A comparative example on 2013 January 4. The data were all acquired with CCD cameras. From left to right in (a), the pre-processed images of BBSO, HSOS, and KSO were taken at 17:11:40 UT, 03:44:13 UT, and 09:17:46 UT on 2013 January 4, respectively. In panel (b), we show histograms of the above images with curve means are 2.1873e+04, 2.1872e+04, and 2.7196e+04, and standard deviations are 1.8320e+03, 1.7132e+03, and 2.1626e+03, respectively. In panel (c), we show their recognition results.}
    \label{fig:filaExtResultOthers}
\end{figure}

\begin{figure}[!htbp]
    \centering
    \includegraphics[width=0.48\textwidth]{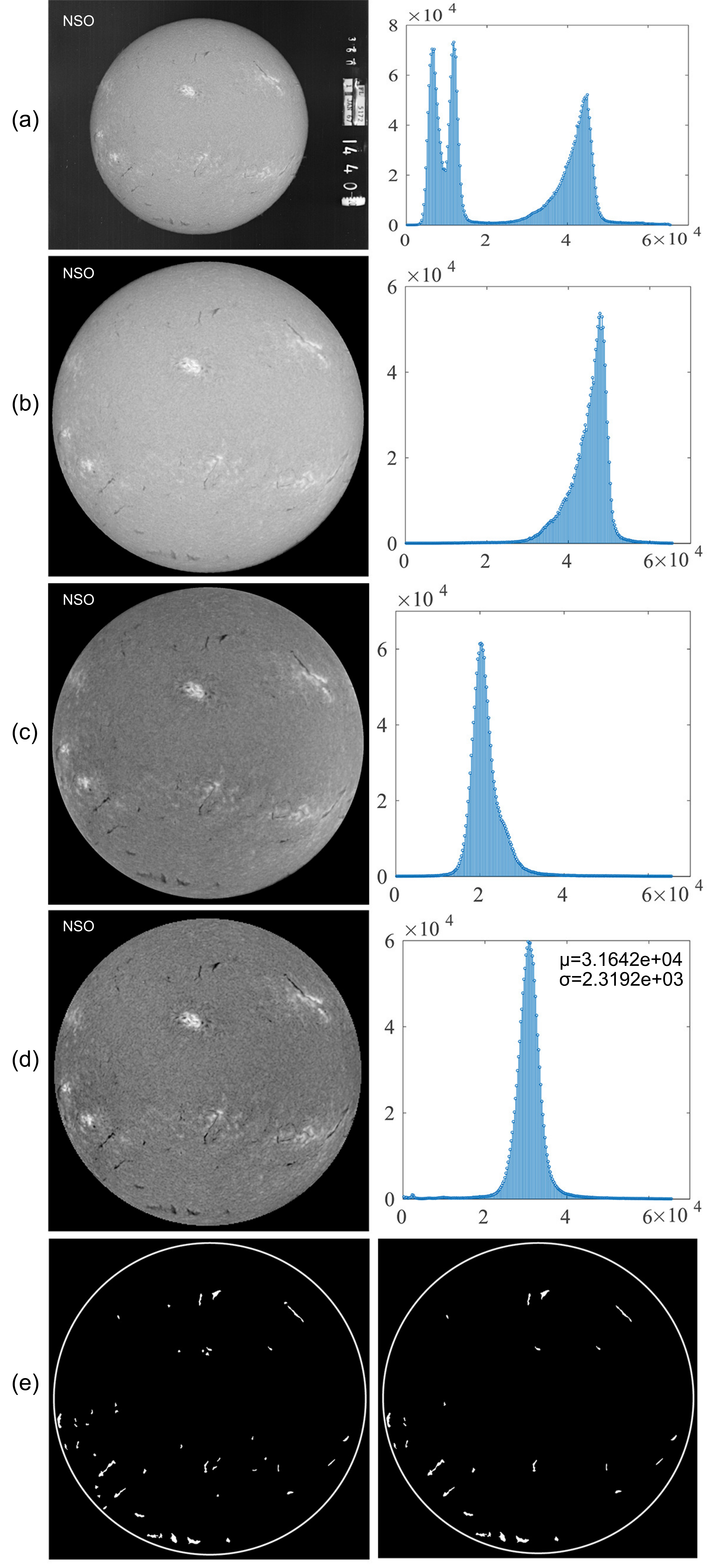}
    \caption{An NSO example at 14:40:04 on 1967 January 1. Its data were originally recorded orignally in photographic plates/films and scanned into a digital image. (a) The plate/film image and its statistical histogram; the same is shown for the panels of (a) -- (d). (b) The digitalized image with a $P$-angle correction. (c) The image with limb darkening being further removed. (d) The image with an uneven background being removed as well and its histogram showing a normal distribution, in which the curve mean is 3.1642e+04 and the standard deviation is 2.3192e+03. (e) Left panel: the recognition result; right panel: the recognition result with the noisy points being removed.}
    \label{fig:filaExtResultNSO}
\end{figure}

This method can recognize and segment filaments of all full disk H$\alpha$ images, obtained from different stations and in different time periods. Figure~\ref{fig:filaExtResult} gives an example of the recognition result. Furthermore, the dynamic threshold we generate can be obtained from every H$\alpha$ image, and it is obtained one time without the need to do looping. Figure~\ref{fig:filaExtResultOthers} shows a comparative example. The data come from three different stations at the same date --- namely BBSO, HSOS, and KSO --- and they all use charge-coupled devices (CCDs) to acquire images during that period. From the figure, we can see that the proposed method can effectively perform filament recognition to the images from different stations at the same date. Figure~\ref{fig:filaExtResultNSO} shows an example of using a digitalized image (originally recorded in photographic plates/films). The proposed method can also perform filament recognition effectively. This algorithm largely reduces the processing time needed. Detailed analyses and demonstration of the algorithm will be presented in a future paper.

After these steps, the filaments can be successfully segmented from the H$\alpha$ image.

However, there are still a certain number of small noise points in the resulting segmentation images, which is an unavoidable disadvantage of using a traditional image processing method. We set a threshold to remove the smaller connected domain. At the same time, we also removed the sunspots by calculating the circularity and the area of each connected region in the binary map. In addition, we also need to calculate the average value of the region in the original image, and those with lower values (dark black) will be removed.

\subsubsection{Parameters of the Filament Feature}

After the above processing and recognition operations, we calculate the various characteristic parameters of filaments, such as the area, perimeter, latitude, longitude, tilt angle, length of filament skeleton, average width of filament, angle between barb and filament skeleton, distance from the starting point to the end point of the filament, and parameters that we believe are useful for a good understanding of the filaments through multi-cycle statistical studies.

\section{Data Products and Query Tool}
\label{sec:dataProd}

\subsection{Products}

The root site of our H$\alpha$ archive is at \url{http://sun.bao.ac.cn/solarfilament/}. In this data archive, we supply the level 0 and level 1 data, binary images, enhanced images, and some of the data products such as animations, synoptic charts, and filament parameter information.

\subsubsection{FITS Data and Intermediate Images}

Level 0 and level 1 FITS data have been described in Section~\ref{sec:dataProc}. During the process of filament recognition, we obtain intermediate images, i.e. enhanced images (as the one in Figure~\ref{fig:filaExtResult}a) and binary images (as the one in Figure~\ref{fig:filaExtResult}d). Enhanced images make the feature recognition much easier, and binary images show the recognition results.

\subsubsection{Animations}

Several types of animations are produced for different targets. Through a visual view of these full disk animations, eruptive events such as flares and filament activities can be easily identified. These movies actually assist us to confirm that the previously wrong orientations in some of the scanned images have been correctly corrected, which is also a tedious procedure that is essential to guaranteeing the correctness of the filament parameters we obtained.

\subsubsection{Synoptic H$\alpha$ Maps}

\begin{figure}[!htbp]
    \centering
    \includegraphics[width=0.75\textwidth]{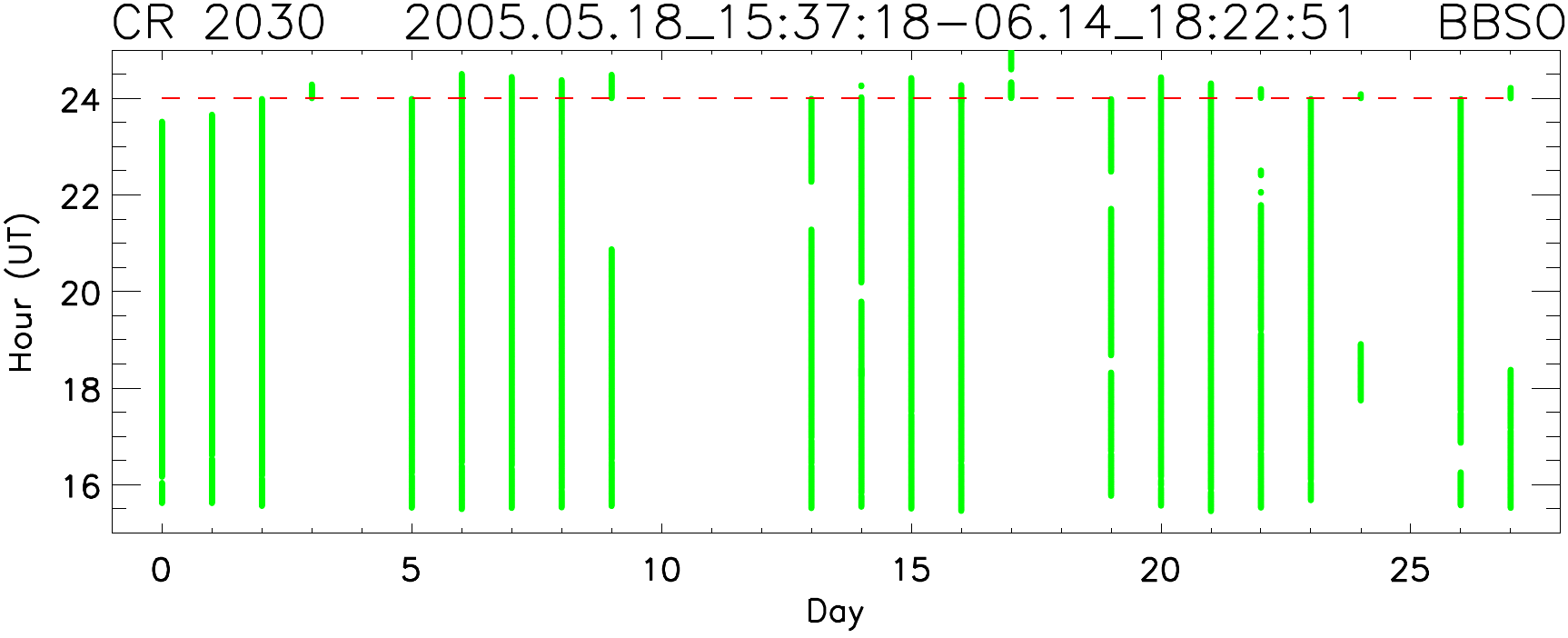}
    \caption{Distribution of BBSO H$\alpha$ observation times in CR2030. The red dashed line corresponds to the hour we selected, in which the Sun has been observed for most days in this CR.}
    \label{fig:CR2030}
\end{figure}

Unlike the space-born solar telescopes, ground-based telescopes cannot observe the Sun all the time, which is often affected by local weather and circumstance. So the H$\alpha$ data from the ground-based observatories are often not continued in time and it is usually difficult to keep a fixed time interval when constructing a H$\alpha$ synoptic map. For a better construction, for each Carrington rotation (CR), we first find out the hour in which the Sun has been observed for most days (see Figure~\ref{fig:CR2030}). Then we chose the best image in each day that is during or closest to that hour.

\begin{figure}[!htbp]
    \centering
    \includegraphics[width=0.9\textwidth]{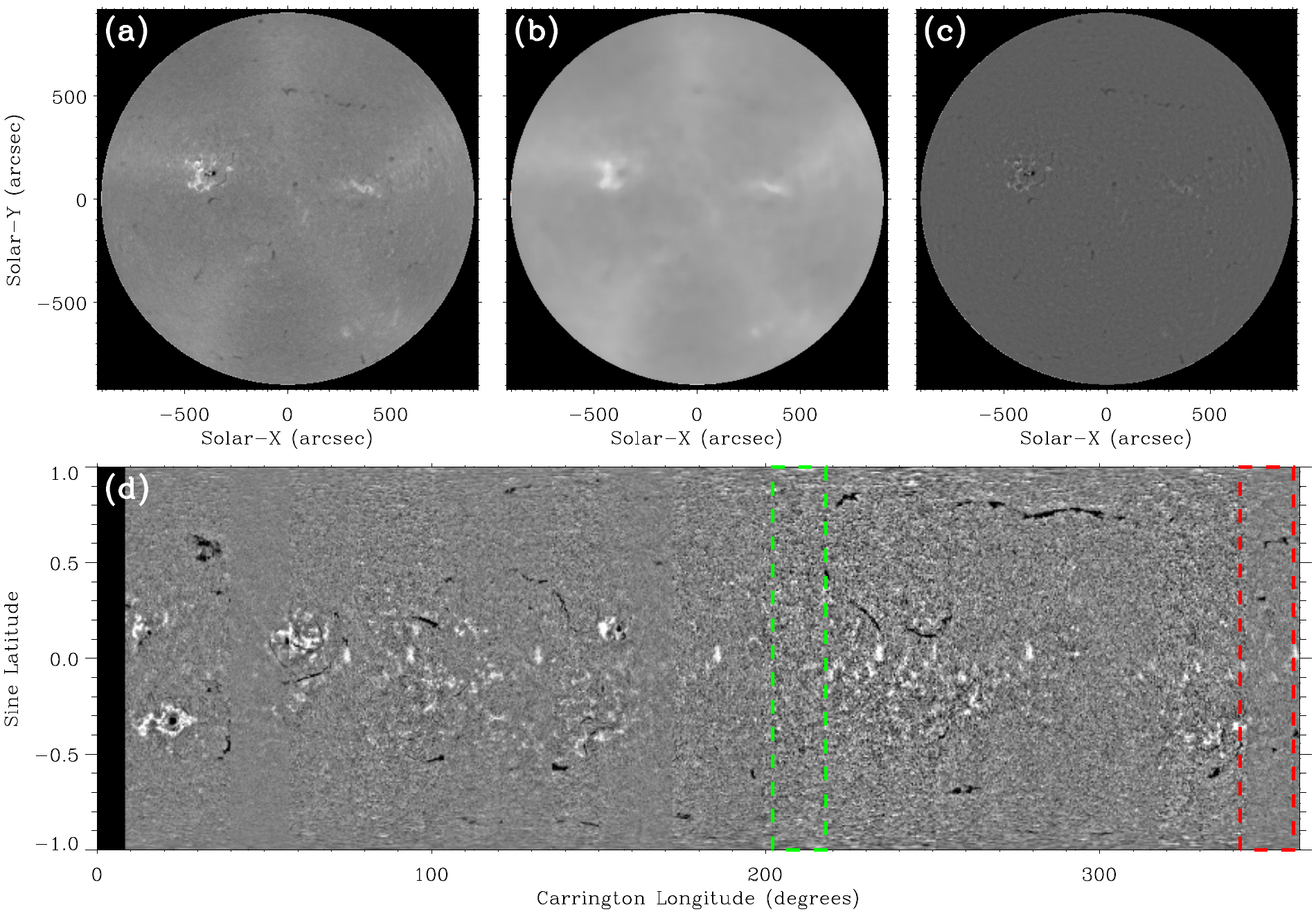}
    \caption{Processing step I. (a) -- (c) An H$\alpha$ image observed by BBSO. Panel (a) is the original image. Panel (b) is the blurred image obtained by applying a large window median filtering on panel (a). Panel (c) is the clean image obtained by subtracting panel (b) from panel (a), which is used to construct the synoptic map. Panel (d) is a the synoptic map CR2032, constructed directly using clean images from the observations of BBSO. Green and red dashed boxes correspond two regions, in which the background intensities are slightly different.}
    \label{fig:synopStep1}
\end{figure}

Before we make the synoptic maps, all the H$\alpha$ data from different observatories have been standardized by correcting the $P$ angle, centering the Sun image, removing the limb darkening, and resizing to $2000\times 2000$ pixels. However, we found that many historical H$\alpha$ data did not deduct the flat field. For example, in Figure~\ref{fig:synopStep1}a, we can see a five claws-like bright structure in this BBSO H$\alpha$ image, which is probably produced by the instrument. Then, as a first step, we obtain the background variation by applying a $50\times 50$ pixel window median filter \citep{Zharkova2005,Bertello2010,Chatterjee2016} on each H$\alpha$ image (Figure~\ref{fig:synopStep1}b). Then we get a clean H$\alpha$ image (Figure~\ref{fig:synopStep1}c) after subtracting the blurred background image.

\begin{figure}[!htbp]
    \centering
    \includegraphics[width=0.9\textwidth]{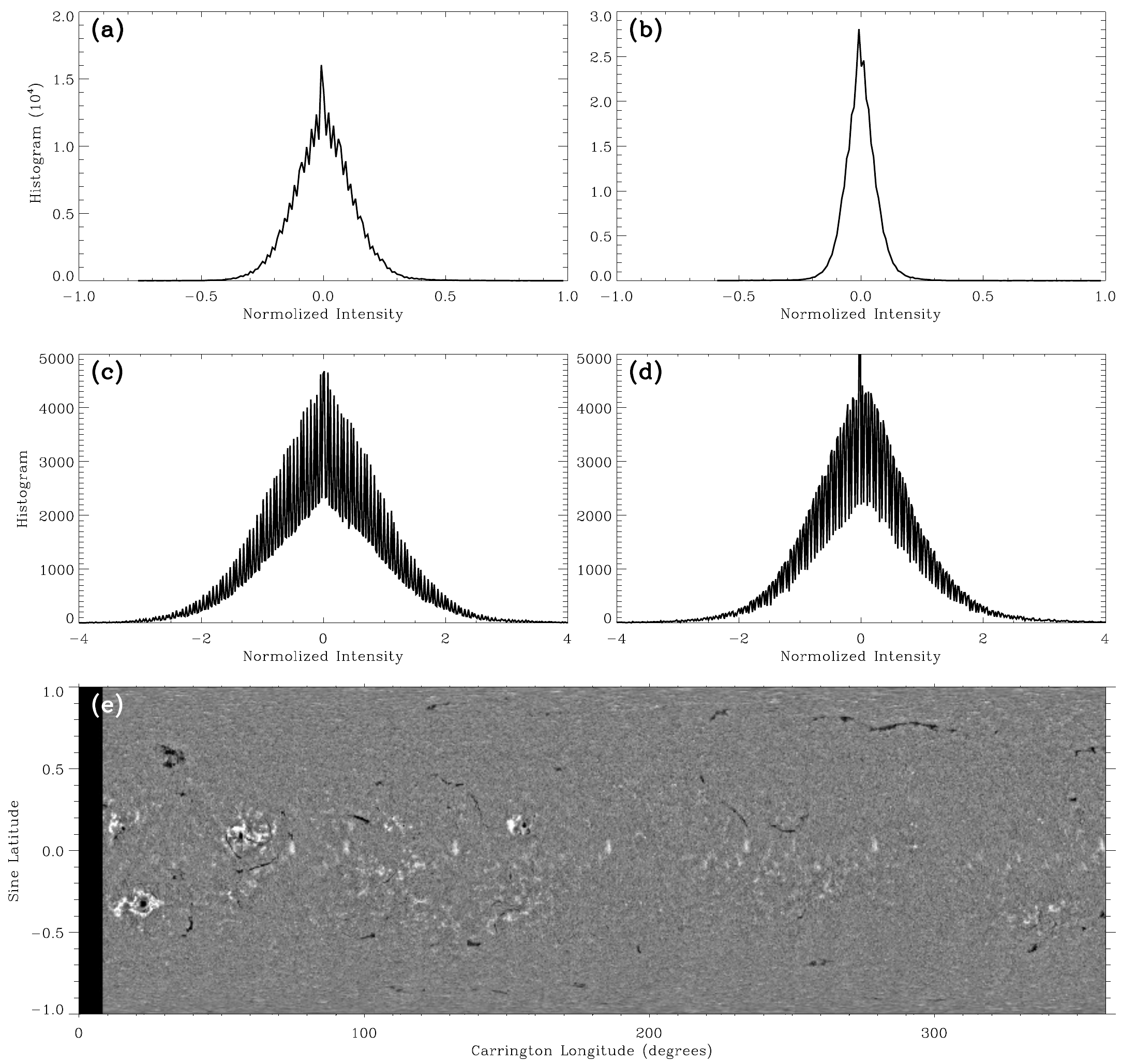}
    \caption{Processing step II. Panels (a) and (b) are the histograms of the normalized intensities in the green and red dashed boxes in Fig.\ref{fig:synopStep1}d. Panels (c) and (d) are the histograms of the normalized intensities in panels (a) and (b) that are divided by $\sigma$ of the corresponding normalized H$\alpha$ intensity image. Panel (e) is the synoptic map CR 2032 (Fig.\ref{fig:synopStep1}d) after the processing step II.}
    \label{fig:synopStep2}
\end{figure}

Following the method introduced by \citet{Harvey1998} and \citet{Ulrich2006}, we construct the H$\alpha$ Carrington map or synoptic map (Figure~\ref{fig:synopStep1}d), with a longitude range of $60^\circ$ ($-30^\circ$ -- $30^\circ$) in each daily image \citep{Sheeley2011,Chatterjee2016,Chatterjee2017}. However, we find that the background intensities for different parts in the Carrington map are slightly different, such as the regions marked by green and red dashed boxes in Figure~\ref{fig:synopStep1}d, though we have already normalized the intensity in each image. Figure~\ref{fig:synopStep2}a and b show the histograms of the normalized intensities in these two regions, which present two different Gaussian shapes: one is wide and short and the other is narrow and tall. This may be caused by the drift of observational lines or variation of seeing in different times.

\begin{figure}[!htbp]
    \centering
    \includegraphics[width=0.7\textwidth]{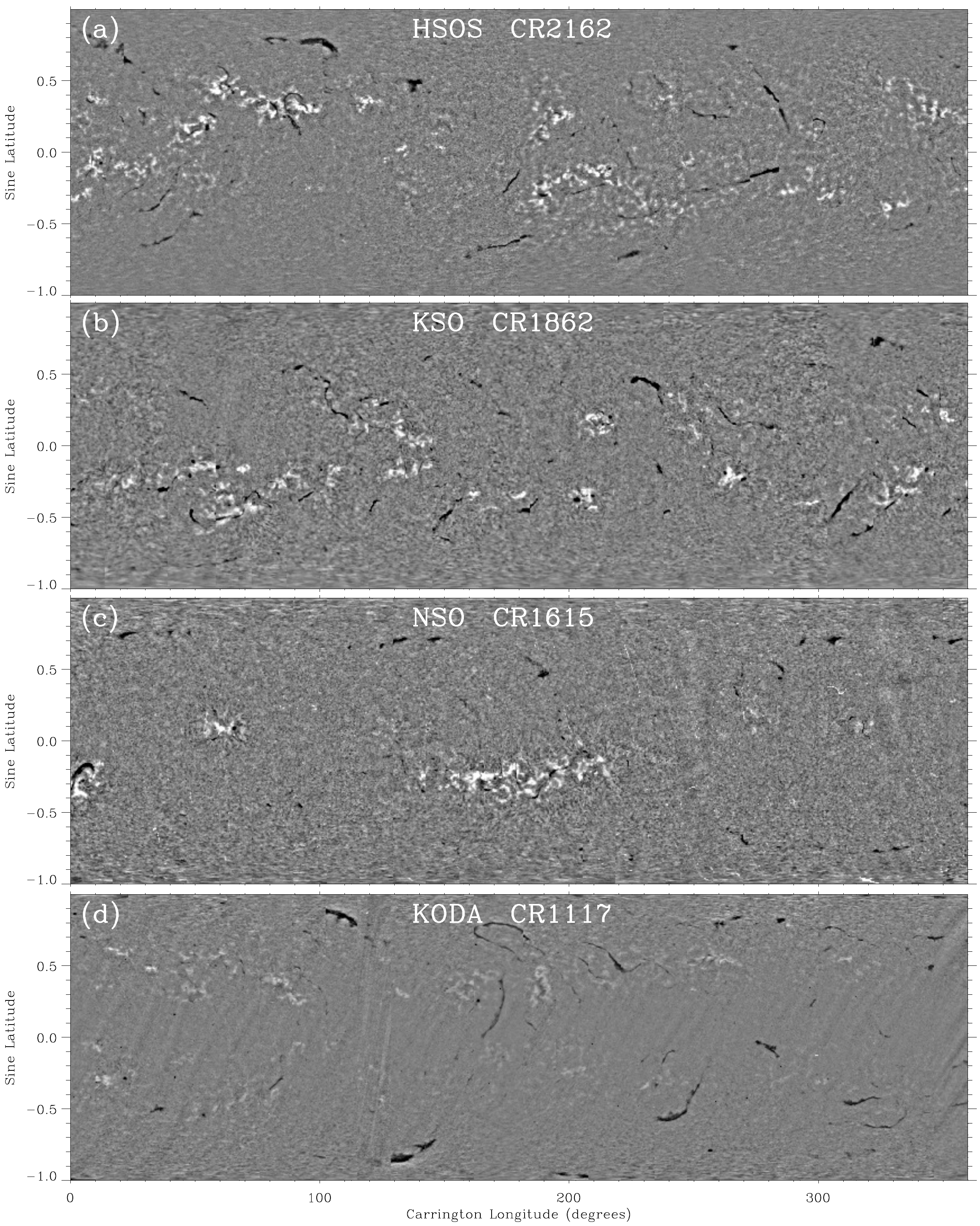}
    \caption{H$\alpha$ synoptic maps, using HSOS, KSO, NSO, and KODA data, respectively.}
    \label{fig:synopMap}
\end{figure}

As the second step, we adjust the distributions of the intensity values of different images to a similar level. We calculated the $\sigma$ value of each normalized H$\alpha$ image. Different images usually have different values of $\sigma$. We make the first normalized image by dividing by $\sigma$, and then we see the distributions of the intensity values in different images are similar (Figure~\ref{fig:synopStep2}c and d). Using the new normalized images, we reconstruct the Carrington map (Figure~\ref{fig:synopStep2}e), which seems much better. The background intensities in the new Carrington map are distributed uniformly. We also apply this method to construct synoptic maps using the H$\alpha$ images from other ground-based solar observatories (Figure~\ref{fig:synopMap}). We can see that they all look reasonably good.

It should be noted that some Carrington maps still contain defects, like small jumps in intensity or discontinuity in chromospheric structures. They may be caused by clouds, bad seeing, and the observations of far off the line center or resulted from the evolution of an active filament or a flare eruption. In the future, we may remove the off line center images and select some more high-quality images in each day to construct a Carrington map.

\subsubsection{Filament Parameters}

Based on the recognized filaments (Section~\ref{sec:filRec}), some of the filament parameters can be calculated. These filament parameters are the longitude and latitude of the filament center, area, perimeter, length and tilt angle of the filament bone, average width, Euclidean distance from head to tail points, average intensity after limb darkening removed, etc. In addition to the date, time, and solar radius at the specific observing moment, a filament information database is constructed, and we also numbered all the filaments subtracted from the full disk images. A text version of the filament information lists can be found in the directory of ``filInfo''.

\subsection{Query Tool}

For a more convenient and efficient use of these feature parameters, a parameter query tool is developed. It provides functions to do parameters searching, visualization, and statistics as well as downloading. The queryable parameters include all features extracted.

\begin{figure}[!htbp]
    \centering
    \includegraphics[width=\textwidth]{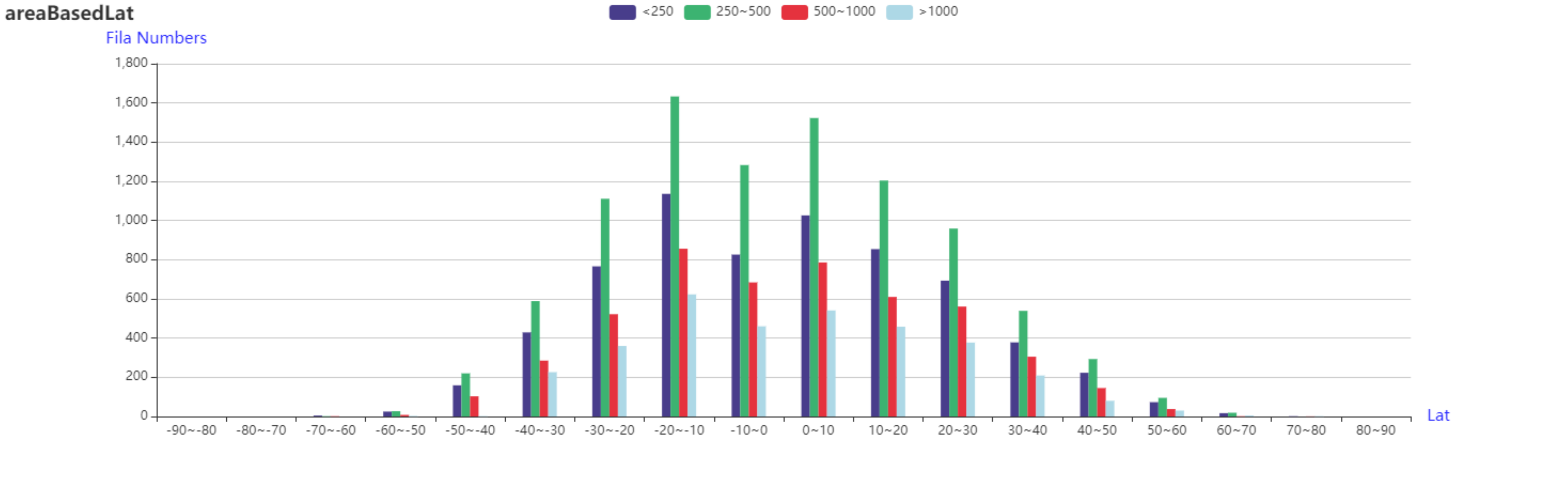}
    \caption{Statistics on filament area in different latitude.}
    \label{fig:areaStat}
\end{figure}

Figure~\ref{fig:areaStat} gives an example of a query statistical result, e.g., the statistical result of our query tool on filament area at different latitudes in solar cycle 22 is compared with the statistical result from \citet{HaoQi2015} and \citet{HaoQ2015}. We can see the two are consistent with each other.

\section{Solar Physics Research Based on the Data Set: An example}

Our dataset can be used to attack many scientific problems. Here we just give an example.

The study of polar crown filaments (PCFs) is very important for understanding the variations of the solar cycles (e.g. \citeauthor{Brajsa1990}~\citeyear{Brajsa1990}), such as the prolonged cycle 23 and the nearly ending ``abnormal'' cycle 24. \citet{XuY2018} statistically investigated the properties of PCFs from 1973 to 2018, such as their move trends at the beginning of each solar cycle, the variations of polar field strength when the PCFs approach to the maximum latitude, and the migration rates for different Carrington rotations in different solar cycles. Further studies are still needed if we want to obtain a deep understanding of the solar cycles.

\section{Summary and Planned Future Work}
\label{sec:conc}

We collected the original H$\alpha$ data from five observatories around the world and merged them into a single dataset to provide the international scientific community with a comprehensive H$\alpha$ dataset covering data over 100 yr. To construct this data set, we have carried out and finished following tasks. All data have been standardized. $P$ angles, possibly aroused in the process of digitalization, are corrected. Time stamp information of the historical 40 yr NSO data has been extracted. Methods of automatically pinpointing the Sun's edge and classifying data by quality have been proposed and applied to all data. A method of an automatic dynamic threshold has been proposed and applied to multiple observatory data. Our dataset has following characters that make it stand out from others: it is more continuous in time sequence; higher in time frequency, especially for the recent 60 yr; and provides eight physical parameters of solar filaments and seven data products, including H$\alpha$ movies since 1912 and synoptic H$\alpha$ maps of all the Carrington rotations since 1912. Meanwhile, a query tool with a visualization function is also provided for querying our filament parameters.

In the future, we plan to further increase the density and the continuity of the data set. This will help the study of filament dynamics, such as the formation and disappearance of filaments.  We also plan to fuse the H$\alpha$ data of the Kislovodsk Mountain Astronomical Station into our dataset, after their digitization work is finished. Further classification of filaments is also planned, such as classifying the filaments according to their position scheme of \citet{Engvold2015} --- for example, intermediate (IP), active region (AR), quiescent (QS) --- which may help the study of the global magnetic topology structure.

\acknowledgements
The authors appreciate the anonymous referee's instructive comments and suggestions to improve the manuscript. We are grateful to the NSO of USA, BBSO of USA, KSO of Austria, and KODA of India for providing the original data. We thank PhD students Quan Wang and Wei Wu for their help checking the $P$ angle in the NSO and KODA data. The research is supported by the National Natural Science Foundation of China under grants U1531247, 11427901, 11803002, and 11973056; the special foundation work of the Ministry of Science and Technology of China under grant 2014FY120300; the 13th Five-year Informatization Plan of Chinese Academy of Sciences under grant XXH13505-04; and US NSF under grant AGS-1620875 and AGS-1928265.


\end{document}